\documentclass[%
 reprint,
 amsmath,amssymb,
 aps,
]{revtex4-2}
\usepackage{graphicx}
\usepackage{dcolumn}
\usepackage{bm}

\usepackage{graphicx}
\usepackage{amsmath}  
\usepackage{hyphenat} 

\usepackage[utf8]{inputenc}
\usepackage{graphicx}
\usepackage{dcolumn}
\usepackage{bm}
\usepackage{float}
\usepackage{amsmath}
\usepackage{graphicx,subfigure,epsfig}
\usepackage{hyperref}
\hypersetup{
    colorlinks=true,
    citecolor=blue,
    linkcolor=blue,
    filecolor=magenta,
    urlcolor=cyan,}

\begin{document}

\preprint{APS/123-QED}

\title{Quasinormal Modes of Schwarzschild Black Holes in the Dehnen-(1, 4, 5/2) Type Dark Matter Halos}

\author{Qi-Qi Liang}
\affiliation{College of Physics, Guizhou University, Guiyang, 550025, China}
\author{Dong Liu}
\affiliation{ Department of Physics, Guizhou Minzu University, Guiyang 550025, China}
\author{Zheng-Wen Long}
\email{zwlong@gzu.edu.cn (Corresponding author)}
\affiliation{College of Physics, Guizhou University, Guiyang, 550025, China}

\date{\today}

\begin{abstract}
The Dehnen - type dark matter density distribution model is mainly used for dwarf galaxies. In recent years, researchers have speculated that black holes may exist in this dark matter model and have given the black hole metric solutions. On this basis, this paper conducts a systematic study on the quasinormal modes of a Schwarzschild black hole in a Dehnen - (1,4, 5/2) dark matter halo, revealing the influences of dark matter distribution and perturbation field types on the black hole's quasinormal modes.The research uses the shadow radius data of the M87$^{\ast}$ black hole. Through the geodesic equation, two sets of dark matter halo parameter values of $\rho_{\rm s}$ and $r_{\rm s}$ are determined, and the specific numerical values of the black hole's event horizon radius, photon sphere radius, and shadow radius under the corresponding conditions are obtained. The wave equations and effective potentials of the black hole under the perturbations of the scalar field, electromagnetic field, and axial gravitational  were analyzed. It was found that the larger the values of $\rho_{\rm s}$ or $r_{\rm s}$, the smaller the peak value of the effective potential, and the wave function oscillation slows down with a lower frequency. The black hole remains stable under perturbations. These studies provide relevant data for the quasinormal modes of the Schwarzschild black hole in the Dehnen-(1,4, 5/2) type dark matter halo. They also offer crucial evidence for understanding the interaction mechanism between the black hole and the dark matter halo.
\end{abstract}

\maketitle

\section{Introduction}

In the exploration of the mysteries of the universe, black holes, as an extremely special and captivating celestial body, have always been a research hotspot regarding their spacetime properties \cite{Allahyari:2019jqz,Shaikh:2018lcc,Sotiriou:2014pfa,Tsukamoto:2017fxq,Wei:2022dzw}. The Schwarzschild black hole, as an idealized black hole model, can theoretically have its spacetime properties described as long as its mass is known \cite{Schwarzschild:1916uq}. However, in the real cosmic environment, such an ideal black hole seems hardly to exist. This is because celestial objects in the universe do not exist in isolation but are constantly interacting with each other. For example, the images of the Virgo A galaxy (M87$^{\ast}$) and Sagittarius A$^{\ast}$ (Sgr A$^{\ast}$)
captured by the Event Horizon Telescope show us the complex structures around black holes \cite{EventHorizonTelescope:2019ths, EventHorizonTelescope:2022wkp}. They both consist of an irregular bright ring structure surrounding a dark central region. This bright ring structure is actually an accretion disk formed by the strong gravitational pull of the black hole attracting matter, which affects the spacetime structure of the black hole. Moreover, the phenomenon of black hole mergers can also lead to the formation of new black holes \cite{Shannon:2015ect}. Beyond that, astronomical observations indicate that, in addition to the baryonic matter familiar to us, which is composed of protons, neutrons, etc., there also exists a vast amount of mysterious dark energy and dark matter in the universe. These substances are subtly influencing the spacetime around black holes \cite{Planck:2015mvg,Green:2020jor, Clesse:2016vqa}. In such a complex and changeable environment, black holes are always in a perturbed state. Under these circumstances, the study of quasinormal modes becomes particularly important. Quasinormal modes can be regarded as a set of precise probes. By analyzing the inherent vibration modes presented by black holes under perturbation, they can accurately capture the subtle changes in key characteristics of black holes, such as mass and spin, in complex environments \cite{Andersson:1996xw, Nollert:1993zz,Nomura:2021efi,Momennia:2019edt}. This provides an indispensable perspective for us to further unlock the mysteries of the complex physical processes around black holes.

Dark matter can be classified in multiple ways. Based on characteristics such as particle properties, motion velocity, and interactions, it is mainly divided into the cold dark matter model \cite{Sollom:2009vd, WMAP:2010qai}, the warm dark matter model \cite{Viel:2005qj,Boyanovsky:2010sv}, the Bose - Einstein condensate dark matter model \cite{Panotopoulos:2017pgv}, and the superfluid dark matter model \cite{Berezhiani:2017tth}.
These different classifications offer a clear framework for researchers to deeply explore the composition of dark matter. They are conducive to a more comprehensive and in - depth understanding of the impact of dark matter on the structure and evolution of the universe. In relevant research fields, numerous scientific research teams have achieved fruitful results  \cite{Xu:2018wow,Liu:2021xfb,Yang:2023tip,Bertone:2018krk,Hochberg:2014kqa}.

With the continuous advancement of research, Dehnen dark matter has gradually come into the spotlight. Since Dehnen proposed the spherically symmetric dark matter density distribution model in 1993, this theory has been increasingly applied in the fields of galactic dynamics and black hole astrophysics \cite{Dehnen:1993uh}. Early studies primarily focused on accurately describing the nuclear bulge density of elliptical galaxies using this model. In 2022, Pantig's team first constructed a black hole model in a dark matter halo - surrounded dwarf galaxy, revealing the modulation effect of dark matter on the black hole's gravitational field \cite{Pantig:2022whj}.
In 2024, Gohain further derived the black hole solution in the Dehnen - (1,4,0) dark matter halo and systematically analyzed its thermodynamic properties and strong - field gravitational characteristics \cite{Gohain:2024eer}. In the same year, Ali et al. found that the strong gravitational lensing effect of black holes in this model is highly consistent with the predictions of general relativity, making it a preferred theory for explaining the peripheral environment of galaxies \cite{Ali:2025rop}. Subsequently, Xamidov and other scholars revealed the constraint mechanism of the dark matter halo on particle orbital dynamics \cite{Xamidov:2025hrj}. 
 Al-Badawi's team was for the first time quantified the correlation between the quasinormal modes of this model and the radius of the photon shadow \cite{Al-Badawi:2024qpt}, and derived the solutions for Schwarzschild-like black holes in the Dehnen-(1, 4, 5/2) dark matter halo \cite{Al-Badawi:2024asn}. Subsequently, Hosseinifar and Alloqulov  have also conducted relevant research on this model \cite{Hosseinifar:2025qtx, Alloqulov:2025ucf}.


 To gain a deeper understanding of the \textbf{S}chwarzschild - like \textbf{B}lack \textbf{H}ole model in a \textbf{D}ehnen-(1,4, 5/2) dark matter halo (\textbf{SBHD}), we conducted a systematic study on its quasinormal modes.  In Section \ref{sec:2}, 
 the ranges of $r_{\rm s}$ and  $\rho_{\rm s}$ are determined based on the actual observational data of M87$^{\ast}$ obtained by the Event Horizon Telescope. Two sets of parameter values are set for subsequent research, and the variations of the effective potential under different field perturbations are investigated.In Section \ref{sec:3}, we analyze the complex frequencies of the quasinormal modes using two numerical methods. The summary is given in Section \ref{sec:4}. In this paper, we mainly adopt the geometric unit system where G=M=
c=1.

\section{ Perturbations of SBHD}
\label{sec:2}
\subsection{Geodesic Equation}
The metric of the Schwarzschild-like black hole in the Dehnen-(1,4,5/2) dark matter halo is given by  \cite{Al-Badawi:2024qpt}:
\begin{equation}
    ds^{2}=-f(r)dt^{2}+\frac{1}{f(r)}dr^{2}+r^{2}(d\theta^{2}+\sin^{2}\theta d\varphi^{2})
\label{equ:1}
\end{equation}

\begin{equation}
    f(r)=1 - \frac{2M}{r}-32\pi\rho_{s}r_{s}^{3}\sqrt{\frac{r + r_{s}}{r_{s}^{2}r}}
 \label{equ:2}   
\end{equation}
M represents the black hole mass, while $\rho_{\rm s}$ and $r_{\rm s}$ denote the core radius and dark matter central density, respectively, characterizing its structure and matter distribution. From the expression of $f(r)$, it is evident that when $ \rho_s = 0 $, the black hole is surrounded by no dark matter, and the spacetime reduces to the Schwarzschild black hole solution.

\begin{figure*}[t!]
  \includegraphics[width=0.48\textwidth]{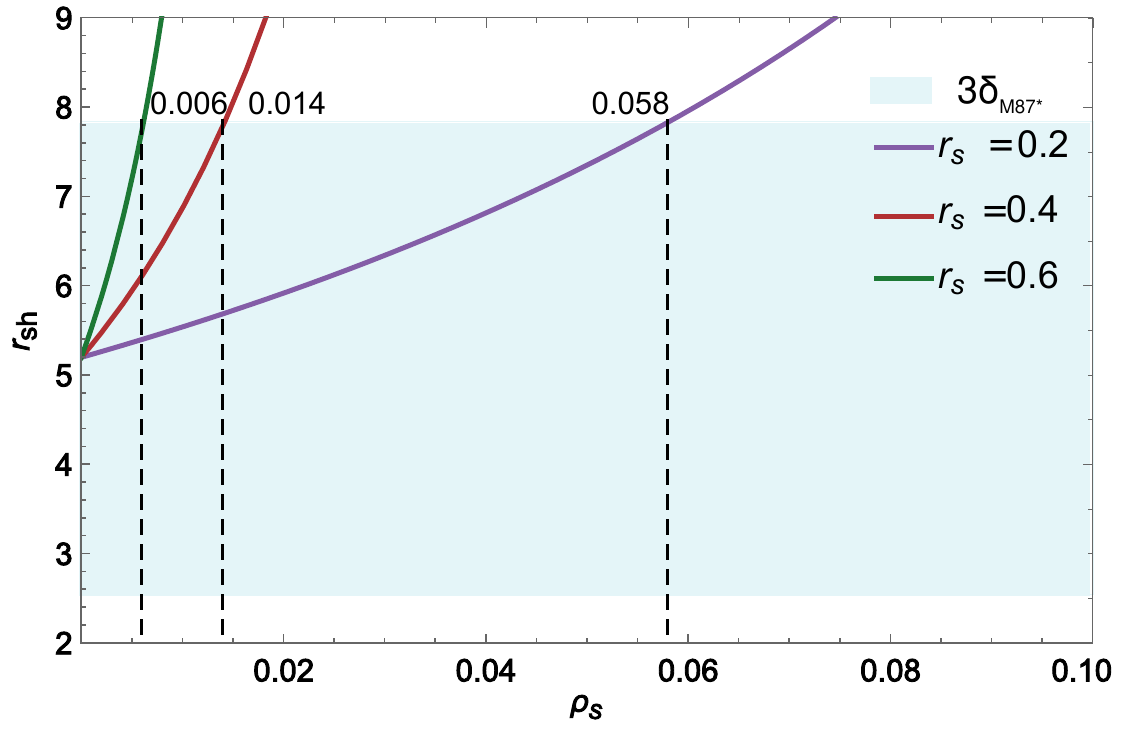}
    \includegraphics[width=0.48\textwidth]{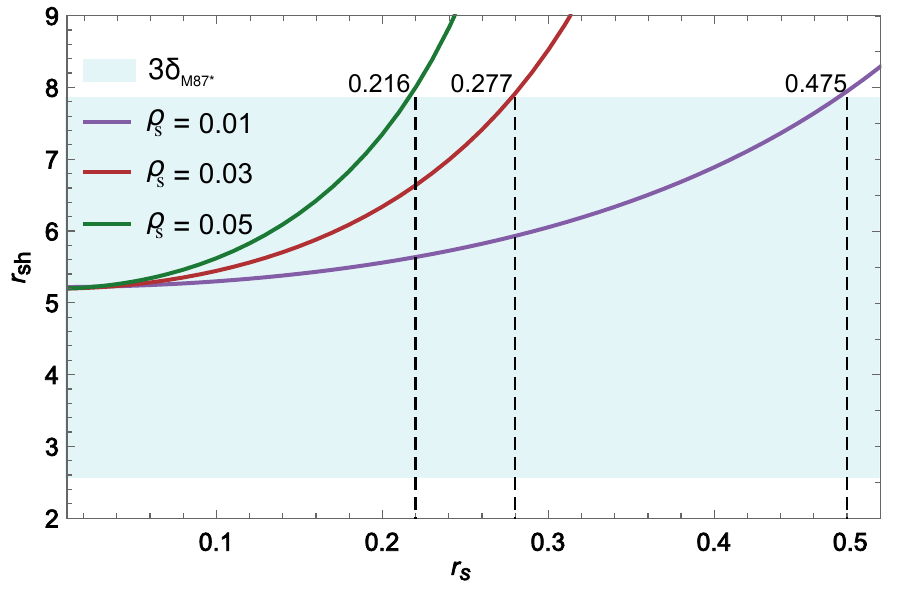}
\caption{The actual observational data of the 3$\delta$ confidence interval of the shadow radius of the M87* black hole constrain $\rho_{\rm s}$ and $r_{\rm s}$, with l = 2.}
\label{fig:1}       
\end{figure*}

It's crucial to determine the values of $\rho_{\rm s}$ and $r_{\rm s}$ prior to conducting the research, as these parameters determine the validity of our subsequent discussions. The geodesic equations of a black hole can be employed to describe the trajectory of photons within the black hole's gravitational field. By analyzing and calculating these trajectories, the black hole shadow can be derived. Utilizing the geodesic equations of this black hole, when considering the equatorial plane $(\theta = \pi/2)$
, the geodesic equations are expressed as:

\begin{multline}
\mathcal{L}=\frac{1}{2}g_{\mu\nu}\frac{dx^{\mu}}{d\lambda}\frac{dx^{\nu}}{d\lambda}\\
=-g_{tt}\left(\frac{dt}{d\lambda}\right)^2+\frac{1}{g_{tt}}\left(\frac{dr}{d\lambda}\right)^2+g_{\varphi\varphi}\left(\frac{d\varphi}{d\lambda}\right)^2
\label{equ:3}
\end{multline}
$\lambda$ is the affine parameter, and the photon energy $E$ and angular momentum $L$ are expressed in comparable terms as:

\begin{equation}
E=-g_{tt}\frac{dt}{d\lambda}=-\left(1 - \frac{2M}{r}-32\pi\rho_s r_s^3\sqrt{\frac{r + r_s}{r_s^2r}}\right)\frac{dt}{d\lambda}
\label{equ:4}
\end{equation}

\begin{equation}
L = g_{\phi\phi} \frac{d\phi}{d\lambda} = r^2 \frac{d\phi}{d\lambda}
\label{equ:5}
\end{equation}
Substituting Equations~\ref{equ:4} and~\ref{equ:5} into~\ref{equ:3}  and considering the the null geodesic case ($\mathcal{L}=0$), we obtain the simplified form:

\begin{equation}
(\frac{dr}{d\lambda})^2 = E^2 - \frac{g_{tt} L^2}{r^2}
\label{equ:6}
\end{equation}
In the above equation, $\frac{g_{tt}}{r^2}$ is defined as the effective quantity $U_{\mathrm{eff}}$, and the photon orbit corresponds to $\left.\frac{dU_{\mathrm{eff}}}{r}\right|_{r_{ph}} = 0$, the $r_{\text{ph}}$  represents the radius of the photon sphere. For a distant observer, the angular radius of the black hole shadow is given by:

\begin{equation}
r_{sh} = \frac{r_{ph}}{\sqrt{1 - \frac{2M}{r_{ph}} - 32 \pi \rho_{s} r_{s}^{3} \sqrt{\frac{r_{ph} + r_{s}}{r_{s}^{2} r_{ph}}}}}
\end{equation}\label{equ:7}

We constrain the parameters using the 3$\delta$ confidence interval of  M87* shadow radius $(2.546M \leq Rs\leq 7.846M)$ \cite{Pantig:2024rmr}, as shown in  Figure~\ref{fig:1}. By comparing the two figures, it can be seen that  $\rho_{\rm s}$ and $r_{\rm s}$ are negatively correlated. When
one of the parameters increases, the other decreases. When $\rho_{\rm s}$ is fixed at $0.01$, the maximum value of $r_{\rm s}$ is $0.475$. When $r_{\rm s}$ is equal to $0.2$, the maximum value of $\rho_{\rm s}$ is $0.058$. 
We select two sets of parameters for subsequent discussions, \( r_s = 0.2 \) with \( \rho_s = 0,  0.01,  0.03,  0.05 \), and \( \rho_s = 0.01 \) with \( r_s = 0.15,  0.3,  0.45 \).

\begin{figure*}[t!]
  \includegraphics[width=0.48\textwidth]{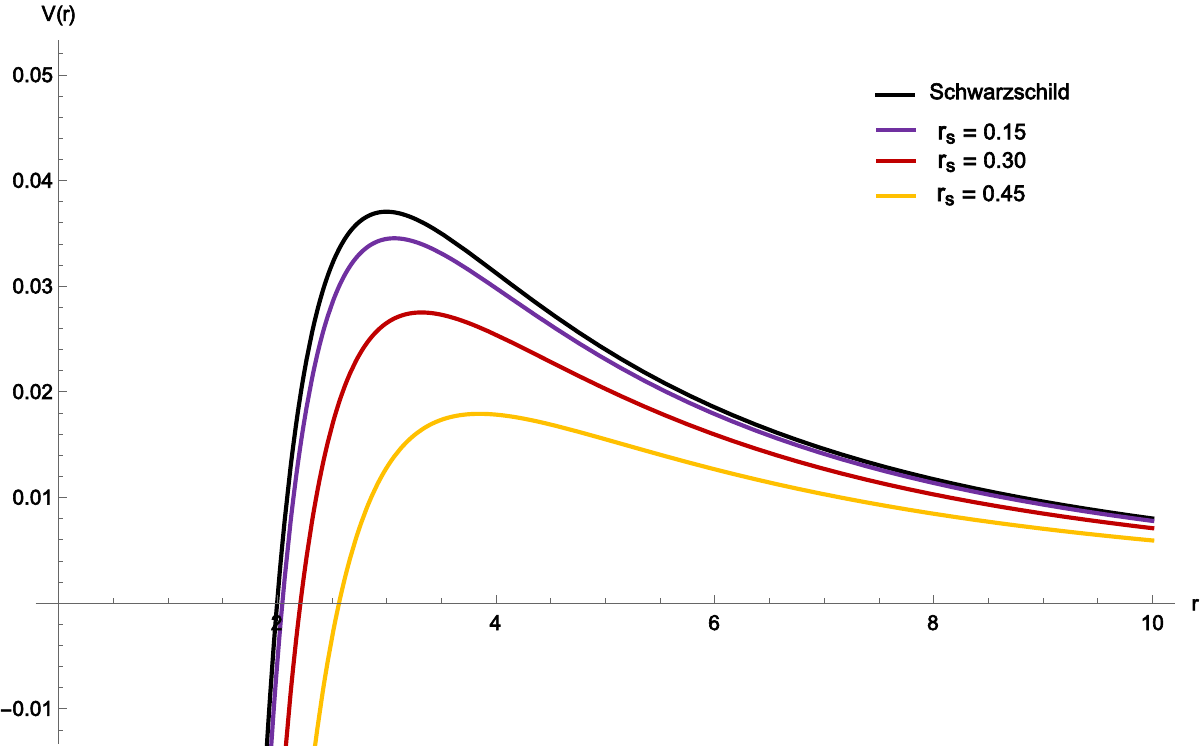}
    \includegraphics[width=0.48\textwidth]{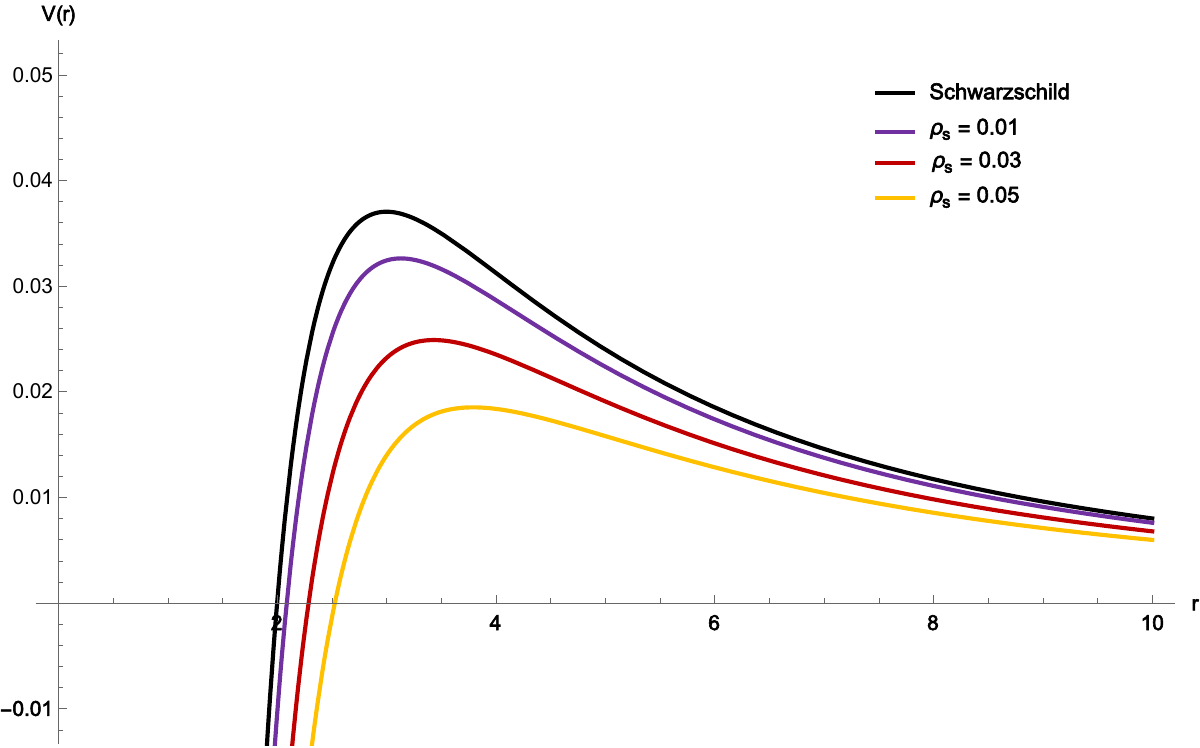}
\caption{The curves of the effective formula U(r) and r of black holes under different parameter values, $ \rho_s = 0.01 $(left), $ r_s = 0.2 $(right), with l = 2.}
\label{fig:2}       
\end{figure*}

To gain a more intuitive understanding of the fundamental spacetime properties of SBHD, we  plotted the effective potential versus radial coordinate r curves in Figure~\ref{fig:2}. According to the definition of the effective potential \(U_{\text{eff}}=\frac{-g_{tt}}{r^{2}}\), the intersection point of the curve with the horizontal axis is the event horizon radius \(r_{\text{h}}\) of the black hole. Moreover, the ordinate of the highest point of the curve corresponds to the value of $\frac{1}{b_{\text{ph}}^2}$ (where \(b = \frac{L}{E}\) is the impact parameter), at this time \(b_{\text{ph}}\) is the shadow radius of the black hole, and the abscissa corresponds to the value of photon sphere radius \(r_{\text{ph}}\). From Figure~\ref{fig:2}, it can be seen that due to the presence of the dark matter halo, both the event  horizon radius and the photon sphere radius of the black hole increase. Moreover, the decrease in the peak value of the effective potential implies an increase in the shadow radius \(b_{\text{ph}}\) of the black hole. In Table \ref{tab:1}, we present the specific numerical values.

\begin{table}
\caption{The corresponding values of each radius under different parameters}
\label{tab:1}       
\begin{tabular}{llll}
\hline\noalign{\smallskip}
 \ & $r_{\text{h}}$ & $r_{\text{ph}}$ & $b_{\text{ph}}$ \\
 \noalign{\smallskip}\hline\noalign{\smallskip}
  Schwarzschild & 2 & 3 & $3\sqrt{3}$ \\
\noalign{\smallskip}\hline\noalign{\smallskip}
$r_{s}$ & \multicolumn{3}{c}{$\rho_{s}=0.01$} \\
\noalign{\smallskip}\hline\noalign{\smallskip}
        0.15 & 2.04799 & 3.07199 & 5.38207 \\
        0.30 & 2.21340 & 3.32018 & 6.03025 \\
        0.45 & 2.56641 & 3.84997 & 7.47359 \\
        \noalign{\smallskip}\hline\noalign{\smallskip}
        $\rho_{s}$ & \multicolumn{3}{c}{$r_{s}=0.2$} \\
        \noalign{\smallskip}\hline\noalign{\smallskip}
        0.01 & 2.08789 & 3.13185 & 5.53704 \\
        0.03 & 2.28781 & 3.43176 & 6.33877 \\
        0.05 & 2.52801 & 3.79209 & 7.34853 \\
\noalign{\smallskip}\hline
\end{tabular}
\end{table}

\subsection{Scalar and Electromagnetic Field Perturbations of  SBHD}
Following the determination of parameter values and fundamental spacetime properties of SBHD, we will proceed to investigate spacetime perturbations around black holes.
The perturbations of the black hole spacetime can generally be classified into external field perturbations and gravitational perturbations. External field perturbations, such as scalar fields and electromagnetic fields, can cause changes in the properties of the spacetime around a black hole when matter carrying these fields approaches the black hole. Gravitational perturbations are divided into two types: axial gravitational perturbations and polar gravitational perturbations. Gravitational perturbations imply that there are subtle changes in the metric of the black hole spacetime, leading to alterations in the degree of spacetime curvature, as well as in the methods of measuring distance and time. In this study, we systematically analyze the scalar field, electromagnetic field, and axial gravitational perturbations of a Schwarzschild-like black hole embedded in a Dehnen-(1,4,5/2) dark matter halo. For the scalar field in curved spacetime, its motion is typically described by the Klein-Gordon equation:

\begin{equation}
    \frac{1}{\sqrt{-g}} \partial_{\mu} (\sqrt{-g} g^{\mu\nu} \partial_{\nu} \Phi) = 0
    \label{eq:example}
\end{equation}
the covariant form of the electromagnetic field's equation of motion is given by:
 
\begin{equation}
\frac{1}{\sqrt{-g}} \partial_{\nu} \left( F_{\rho \sigma} g^{\rho \mu} g^{\sigma \nu} \sqrt{-g} \right) = 0
\label{equ:9}
\end{equation}

\begin{equation}
F_{\rho \sigma}=\partial_{\rho} A^{\sigma}-\partial_{\sigma} A^{\rho}
\end{equation}
Here, $g$ represents the determinant of the metric tensor $g_{\rho\mu}$, $g^{\rho\mu}$ is the contravariant metric tensor of $g_{\rho\mu}$, $\Phi$ is the wave - function of the scalar field, and $A_{\nu}$ is the magnetic four - potential. 

When we substitute the above - mentioned relevant quantities into the equation and consider the spherical symmetry of the metric, the symmetry of the problem is simplified.
Meanwhile, by introducing the tortoise coordinate $\mathrm{d}r_* = \frac{\mathrm{d}r}{f(r)}$, the complex spacetime structure near the black hole event horizon can be transformed to a certain degree, enabling the expressions of some physical quantities to be more concise and easier to handle. Through these steps, a wave equation can be derived:

\begin{equation}
-\frac{d^{2} \Psi}{dr_{*}^{2}} + V(r) \Psi = \omega^{2} \Psi
\label{equ:11}
\end{equation}
Here, $V(r)$ represents the effective potential of the black - hole spacetime when the matter field is perturbed, which plays a crucial role in studying the dynamical behavior of the matter field. For a Schwarzschild black hole surrounded by a Dehnen - (1, 4, 5/2) dark matter halo, the effective potential during scalar - field perturbations is given by:

\begin{multline}
V(r) = \left( 1 - \frac{2M}{r} - 32 \pi \rho_s r_s^3 \sqrt{\frac{r + r_s}{r_s^2 r}} \right)\\
\left[ \frac{l(l + 1)}{r^2} + \frac{2M}{r^3} - \left( \frac{16 \pi \rho_s r_s^3 \left( \frac{1}{r_s^2 r} - \frac{r_s + r}{r_s^2 r^2} \right)}{r \sqrt{\frac{r_s + r}{r_s^2 r}}} \right) \right]
\label{equ:12}
\end{multline}
Here, $l$ is the angular quantum number. The effective potential during electromagnetic - field perturbations is given by:

\begin{equation}
V(r)=\left(1 - \frac{2M}{r}-32\pi\rho_s r_s^3 \sqrt{\frac{r + r_s}{r_s^2 r}}\right)\left(\frac{l(l + 1)}{r^2}\right)
\label{equ:13}
\end{equation}

\subsection{Axial Gravitational Perturbations of SBHD}
The axial perturbation of a black hole refers to the small changes in the spacetime structure or matter distribution of the black hole under the condition of axial symmetry, which can usually be treated with perturbation theory. In this case, the metric $g_{\mu\nu}$ of the black hole can be expressed as the sum of the unperturbed background metric $\bar{g}_{\mu\nu}$ and a small perturbation term $h_{\mu\nu}$:

\begin{equation}
g_{\mu\nu}=\bar{g}_{\mu\nu}+h_{\mu\nu}
\label{equ:14}
\end{equation}
The background metric $\bar{g}_{\mu\nu}$ represents the stable spacetime structure of a black hole when it is not significantly disturbed by external factors. It follows the classical form of the black - hole metric, and for the axial perturbation, there is a perturbation term $h_{\mu\nu}\ll\bar{g}_{\mu\nu}$.

The application of perturbation theory in the axial perturbation of black holes is not limited to the metric. The Christoffel symbols $\Gamma^{\lambda}_{\mu\nu}$ and the Ricci tensor $R_{\mu\nu}$ related to it also need to have respective perturbation terms added under this perturbation situation:
\begin{equation}
\Gamma^{\lambda}_{\mu\nu}=\bar{\Gamma}^{\lambda}_{\mu\nu}+\delta\Gamma^{\lambda}_{\mu\nu}
\label{equ:15}
\end{equation}

\begin{equation}
R_{\mu\nu}=\bar{R}_{\mu\nu}+\delta R_{\mu\nu}
\end{equation}
$\delta\Gamma^{\alpha}_{\mu\nu}$ and $ \delta R_{\mu\nu}$ are expressed as:
\begin{equation}
\delta\Gamma^{\lambda}_{\mu\nu}=\frac{1}{2}\bar{g}^{\lambda\beta}(h_{\mu\beta ;\nu}+h_{\nu\beta ;\mu}-h_{\mu\nu ;\beta})
\end{equation}

\begin{equation}
\delta R_{\mu\nu}=\delta\Gamma^{\lambda}_{\mu\lambda ;\nu}-\delta\Gamma^{\lambda}_{\mu\nu ;\lambda}
\end{equation}
$\bar{\Gamma}^{\lambda}_{\mu\nu}$ and $\bar{R}_{\mu\nu}$ are the Christoffel symbols and the Ricci tensor, respectively, calculated based on the background metric $\bar{g}_{\mu\nu}$. $\delta\Gamma^{\lambda}_{\mu\nu}$ and $\delta R_{\mu\nu}$ are the perturbation terms of the Christoffel symbols and the Ricci tensor, respectively, induced by the metric perturbation term $h_{\mu\nu}$. They reflect the influence of the perturbation on the spacetime geometric connection and the curvature distribution of the black - hole spacetime. Since the perturbation of the background field makes no contribution relative to the background field, it follows that \cite{Kobayashi:2012kh}:
\begin{equation}
\delta R_{\mu\nu} = 0
\end{equation}

In this subsection, we consider the axial gravitational perturbations  that satisfy the Regge - Wheeler (RW)  gauge\cite{Kokkotas:1999bd}. The RW gauge utilizes the characteristics of spherical symmetry and axial perturbations to continuously adjust the constraints on the perturbed metric, thereby obtaining a solvable wave equation that describes the perturbation behavior. At this time, the perturbation term $h_{\mu\nu}$ can be expressed as:

\begin{equation}
h_{\mu\nu} =
\left(
\begin{array}{cccc}
0 & 0 & 0 & h_0(t,r) \\
0 & 0 & 0 & h_1(t,r) \\
0 & 0 & 0 & 0 \\
h_0(t,r) & h_1(t,r) & 0 & 0 \\
\end{array}
\right)
\sin\theta\, \partial\theta P_\ell(\cos\theta)
\end{equation}
Here, $h_0(t,r)$ and $h_1(t,r)$ are functions dependent on time $t$ and radial coordinate $r$, which characterize the specific characteristics of gravitational perturbations at different times and radial positions. $P_l(\cos\theta)$ represents the Legendre polynomial, where $l$ is the order. Substituting Equations~\ref{equ:9} and~\ref{equ:15} into~\ref{equ:14}:

\begin{multline}
\frac{\partial^2 \psi}{\partial t^2} - \frac{f}{r} \frac{\partial}{\partial r} \left[ f \frac{\partial}{\partial r} (r \psi) \right] + \frac{2f^2}{r^2} \frac{\partial}{\partial r} (r \psi) \\+ f \left[ \frac{l(l + 1) - 2f - 2rf'}{r^2} \right] \psi = 0
\end{multline}

From the above equation where $\psi(t,r) = \frac{f(r)}{r} h_1(t,r)$,  introducing the tortoise coordinate $\mathrm{d}r_* = \frac{\mathrm{d}r}{f(r)}$ allows us to derive the wave equation and effective potential for axial gravitational perturbations as follows: 

\begin{equation}
\frac{\partial^2 \psi(t,r)}{\partial t^2} - \frac{\partial^2 \psi(t,r)}{\partial r_*^2} + V(r) \psi(t,r) = 0
\label{equ:22}
\end{equation}

\begin{multline}
V(r) = \left( 1 - \frac{2M}{r} - 32\pi\rho_s r_s^3 \sqrt{\frac{r + r_s}{r_s^2 r}} \right) \\
\left[ \frac{l(l + 1)}{r^2} - \frac{6M}{r^3} + \left( \frac{48\pi\rho_s r_s^3 \left( \frac{1}{r_s^2 r} - \frac{r_s + r}{r_s^2 r^2} \right)}{r \sqrt{\frac{r_s + r}{r_s^2 r}}} \right) \right]
\label{equ:23}
\end{multline}

\begin{figure*}[t!]
    \begin{minipage}[b]{0.32\textwidth}
        \centering
        \includegraphics[width=\textwidth]{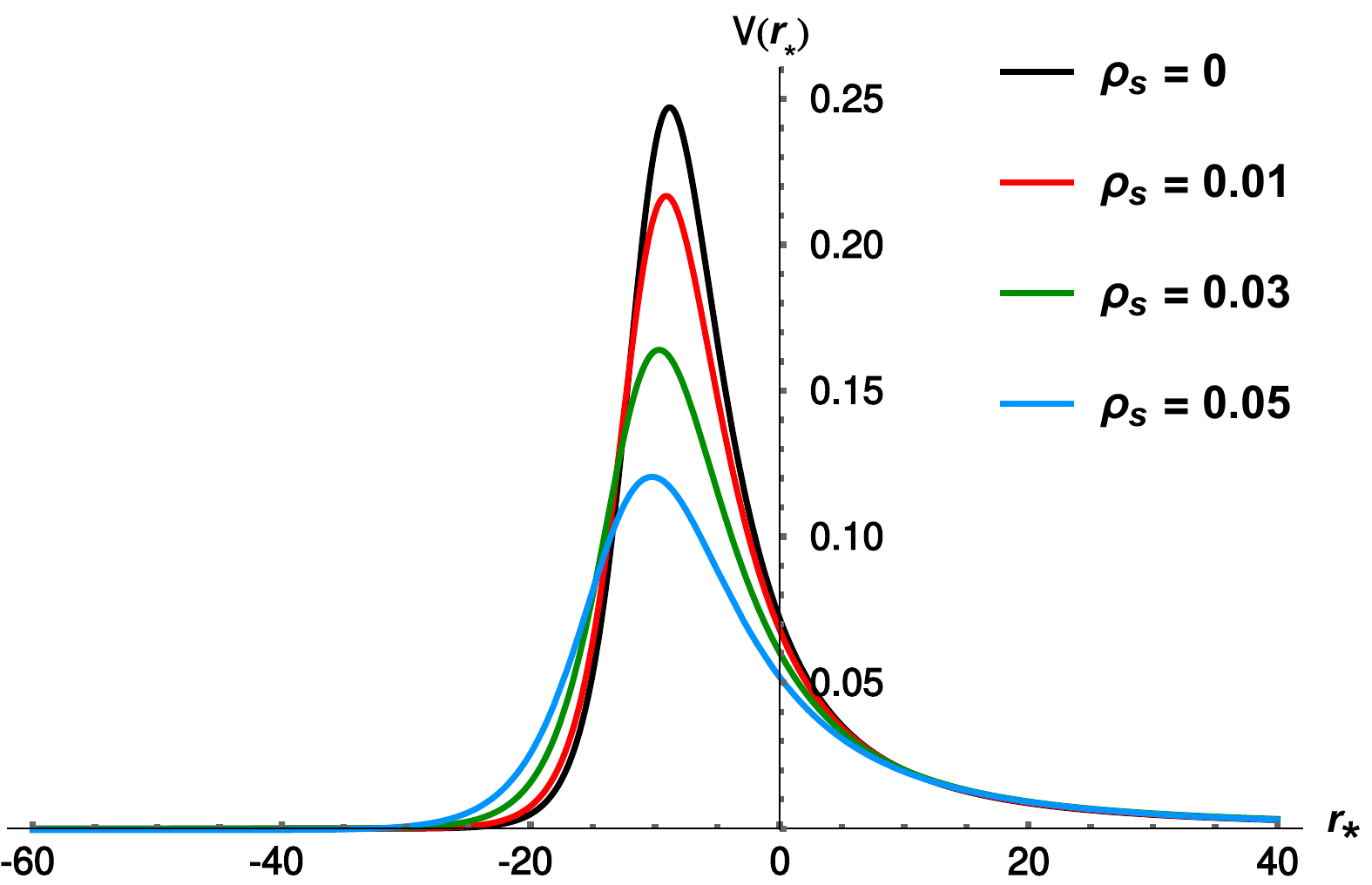}
    \end{minipage}
    \begin{minipage}[b]{0.32\textwidth}
        \centering
        \includegraphics[width=\textwidth]{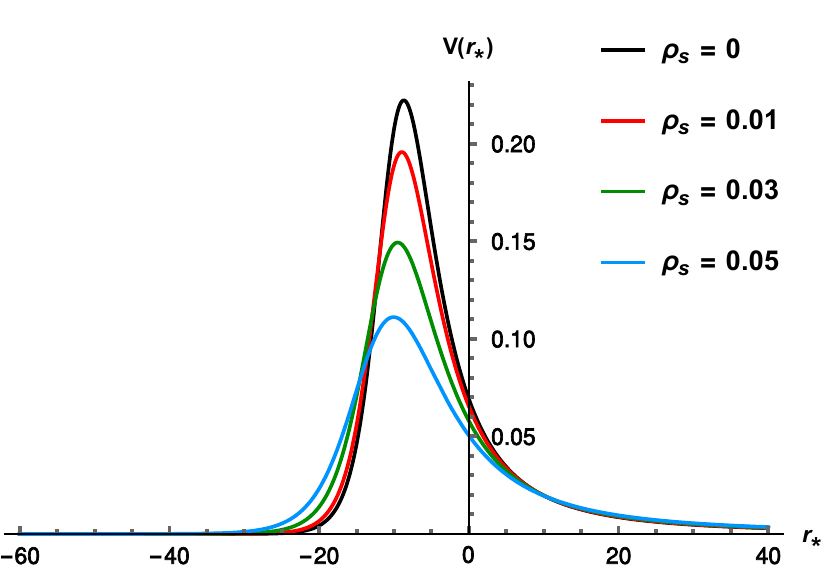}
    \end{minipage}%
    \begin{minipage}[b]{0.32\textwidth}
        \centering
        \includegraphics[width=\textwidth]{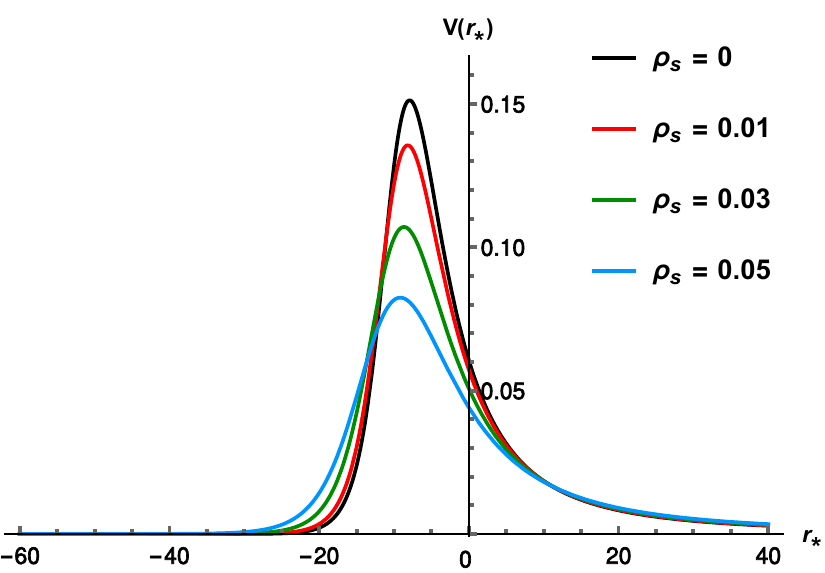}
    \end{minipage}
    \caption{The variations of the effective potential under different perturbations are presented, with the left-to-right sequence corresponding to the scalar field, electromagnetic field, and axial gravitational field ($ r_s = 0.2 $, $ l = 2 $).}
    \label{fig:3}
\end{figure*}

\begin{figure*}[t!]
    \begin{minipage}[b]{0.32\textwidth}
        \centering
        \includegraphics[width=\textwidth]{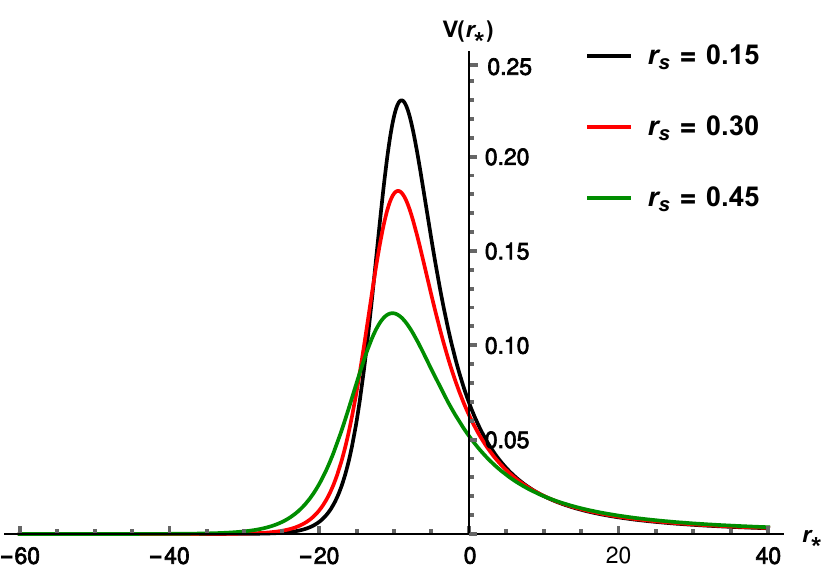}
    \end{minipage}
    \begin{minipage}[b]{0.32\textwidth}
        \centering
        \includegraphics[width=\textwidth]{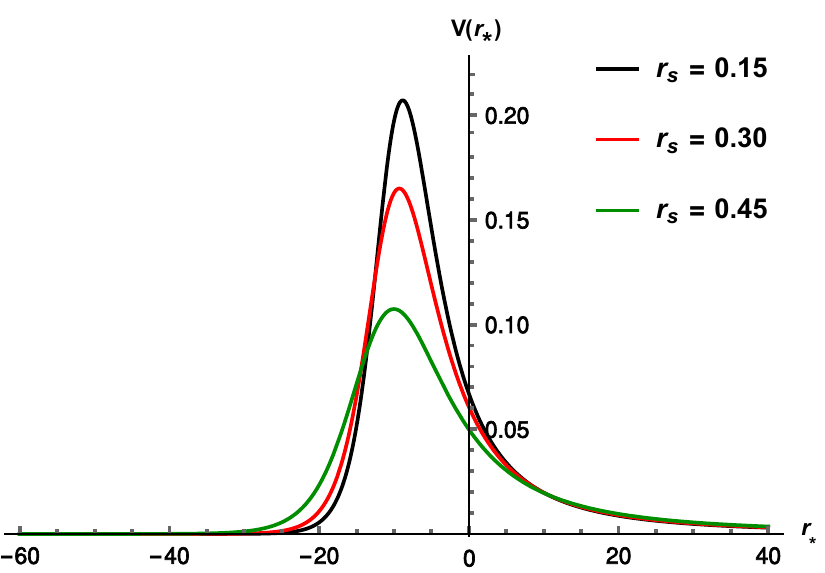}
    \end{minipage}%
    \begin{minipage}[b]{0.32\textwidth}
        \centering
        \includegraphics[width=\textwidth]{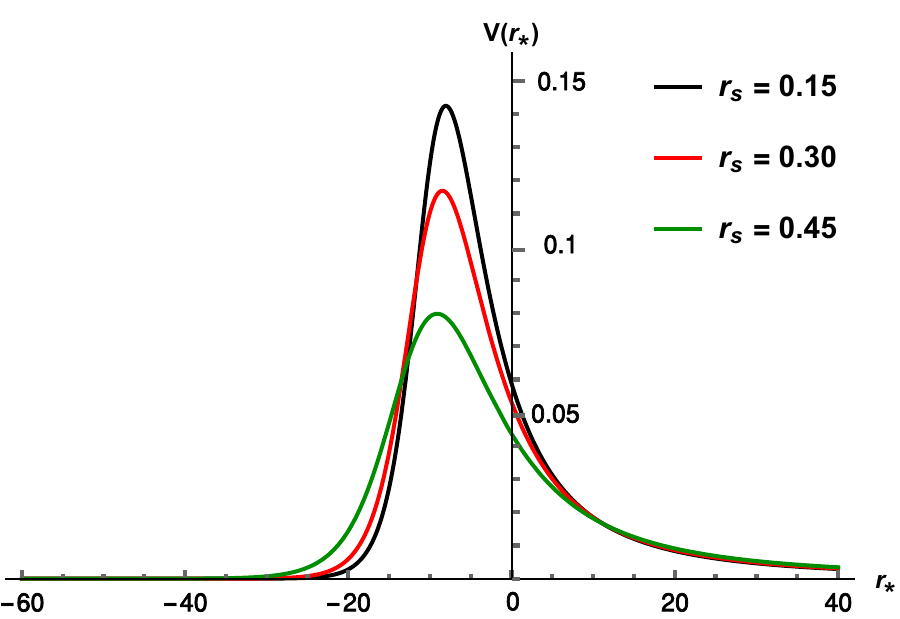}
    \end{minipage}
    \caption{The variations of the effective potential under different perturbations are presented, with the left-to-right sequence corresponding to the scalar field, electromagnetic field, and axial gravitational field ($ \rho_s = 0.01 $, $ l = 2 $).}
    \label{fig:4}
\end{figure*}

Equations~\ref{equ:12},~\ref{equ:13} and~\ref{equ:23} demonstrate that the effective potentials of
different fields exhibit distinct behaviors. The effective potential not only determines the quasinormal modes of black holes but also reflects the particle motion states and system stability, providing critical insights into the physical properties of black holes and related astrophysical phenomena. We investigate the influence of $ r_s $ and $ \rho_s $ on the effective potential through Figure~\ref{fig:3} and~\ref{fig:4}. demonstrate that the presence of dark matter halos reduces the effective potential of black holes. Furthermore, the effective potential peaks diminish with increasing $ r_s $ or $ \rho_s $. $ \rho_s = 0 $ corresponds to the absence of dark matter around the black hole, reducing it to a Schwarzschild black hole. The decrease in the effective potential indicates that the energy "barrier" faced by particles when moving around the black hole is reduced. As a result, particles can move more easily through the space around the black hole, and their activity level increases. Meanwhile, with the decrease in the effective potential, the ``resistance" of the system to external perturbations weakens, making it more vulnerable to external factors that can cause it to deviate from its original state. This promotes changes in the matter distribution and energy transfer around the black hole.

\section{The Quasinormal Modes of SBHD}
\label{sec:3}

When a black hole is perturbed by external disturbances, the spacetime and field distributions around it deviate from their original stable states. The black hole subsequently returns to stability through gravitational wave emission. During this process, the vibrational modes of the fields exhibit specific characteristics, with their corresponding frequencies defined as quasi-normal mode frequencies \cite{Regge:1957td}. Although dark matter remains invisible, it interacts with black holes via gravitational coupling, altering spacetime properties around the black hole and consequently modulating the quasi-normal modes. This section investigates black hole quasi-normal modes 
\cite{Konoplya:2003ii,Iyer:1986nq} using both WKB and time-domain numerical methods.

\subsection{Numerical Methods}
In the previous section, the wave equations of the black hole under perturbations of three kinds of fields were derived. Through the corresponding transformations and considering that the wave equations are independent of time, both Equations ~\ref{equ:11} and~\ref{equ:22} can be simplified to:

\begin{equation}
\frac{d^2 \psi(r_*)}{dr_*^2} + \left[ \omega^2 - V(r_*) \right] \psi(r_*) = 0
\label{equ:24}
\end{equation}

By solving the above equation, under the given appropriate boundary conditions (such as the behavior of the field perturbation at the event horizon of the black hole and at infinity), the specific form of the wave function $\psi(r_*)$ of the field perturbation can be obtained, and then key physical quantities such as the frequency $\omega$ of the field perturbation can be acquired. When solving the above equation, we adopt the sixth-order WKB method. The WKB method can provide an approximate analytical solution. The wave function is expressed as an exponential-form asymptotic expansion:

\begin{equation}
\frac{i(\omega^2 - V_0)}{\sqrt{-2V_0''}} - \sum_{i = 2}^{6} \Lambda_i = n + \frac{1}{2}, \ (n = 0, \, 1, \, 2 \dots) 
\end{equation}

$V_0$ is the maximum value of the effective potential, $\Lambda_i$ is the $i$ order correction term, and $n$ is the overtone number of the quasinormal mode. Different values of $n$ correspond to different vibration modes. When $(n = 0)$, it is the fundamental frequency mode, which is also the mode discussed in this paper.

In addition to using the sixth-order WKB method to find the characteristic frequency when the black hole is perturbed, time -domain methods  can also be adopted. The derivative is approximated by the difference, which is used to discretize the continuous equation. Then, the solution region is divided into grids, and the function values are calculated at the grid nodes In this way, the wave equation can be numerically solved. The light cone coordinates are introduced:

\begin{equation}
u := t - r_* \quad v := t + r_*
\end{equation}
Equations~\ref{equ:11} and ~\ref{equ:22}  can be rewritten as:
\begin{equation}
-4 \frac{\partial^2 \psi(u,v)}{\partial u \partial v} = V \left( \frac{u - v}{2} \right) \psi(u, v)
\label{equ:27}
\end{equation}
To solve equation~\ref{equ:27} , it can be discretized as equation \cite{Gundlach:1993tp}:
\begin{multline}
\psi_{i,j} = \psi_{i,j - 1} + \psi_{i - 1,j} - \psi_{i - 1,j - 1} \\
- \delta^2 \frac{V_{i - 1,j - 1} (\psi_{i,j - 1} + \psi_{i - 1,j})}{8} + \O(\delta^4)
\end{multline}
$\psi_{i,j}$, $\psi_{i,j - 1}$, $\psi_{i - 1,j}$ and $\psi_{i - 1,j - 1}$   correspond to the values of the field perturbation wave function at specific positions, respectively, reflecting information such as the perturbation intensity of the field at these positions. $\delta$ is the grid scale 
\begin{figure*}[t!]
    \begin{minipage}[b]{0.32\textwidth}
        \centering
        \includegraphics[width=\textwidth]{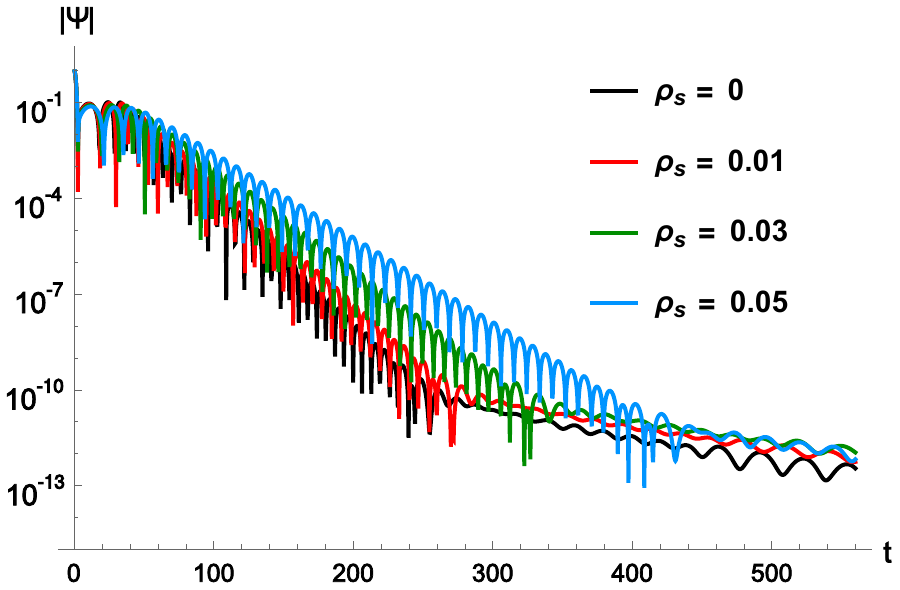}
    \end{minipage}
    \begin{minipage}[b]{0.32\textwidth}
        \centering
        \includegraphics[width=\textwidth]{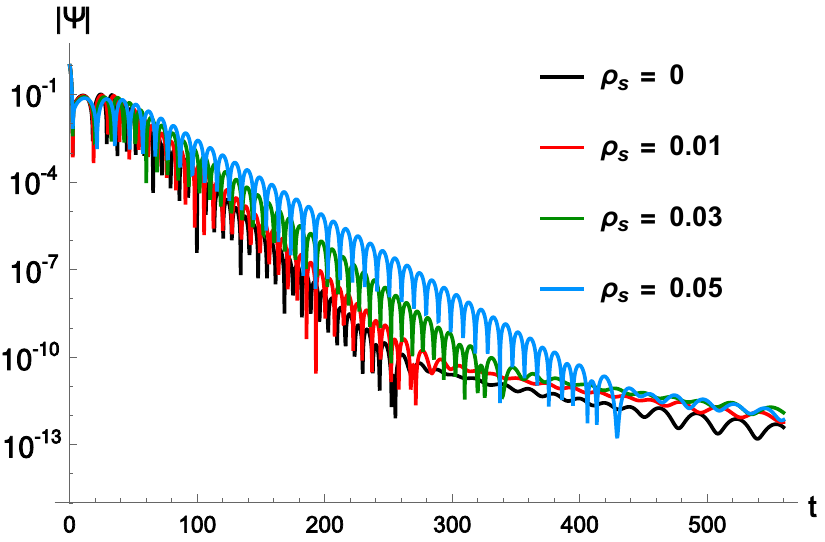}
    \end{minipage}%
    \begin{minipage}[b]{0.32\textwidth}
        \centering
        \includegraphics[width=\textwidth]{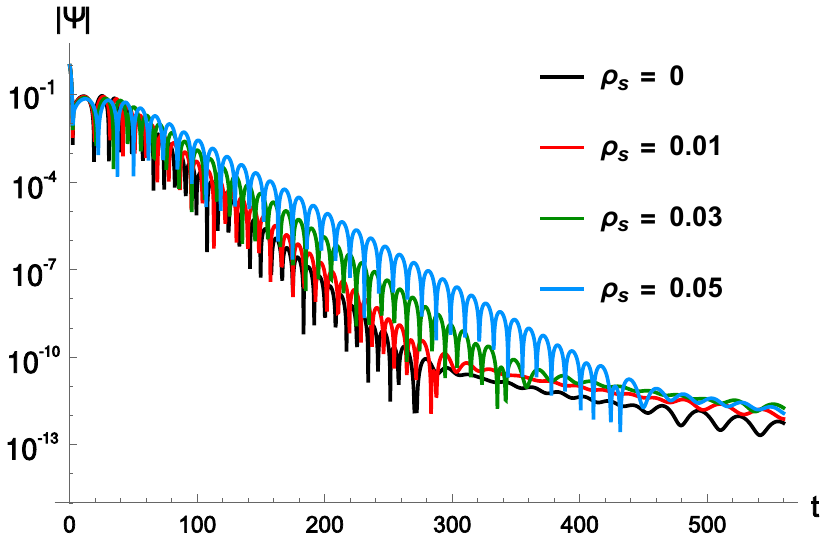}
    \end{minipage}
    \caption{The time-domain profiles from left to right correspond to the scalar field, electromagnetic field, and axial gravitational field  ($ r_s = 0.2 $, $ l = 2 $).}
    \label{fig:5}
\end{figure*}

\begin{figure*}[t!]
    \begin{minipage}[b]{0.32\textwidth}
        \centering
        \includegraphics[width=\textwidth]{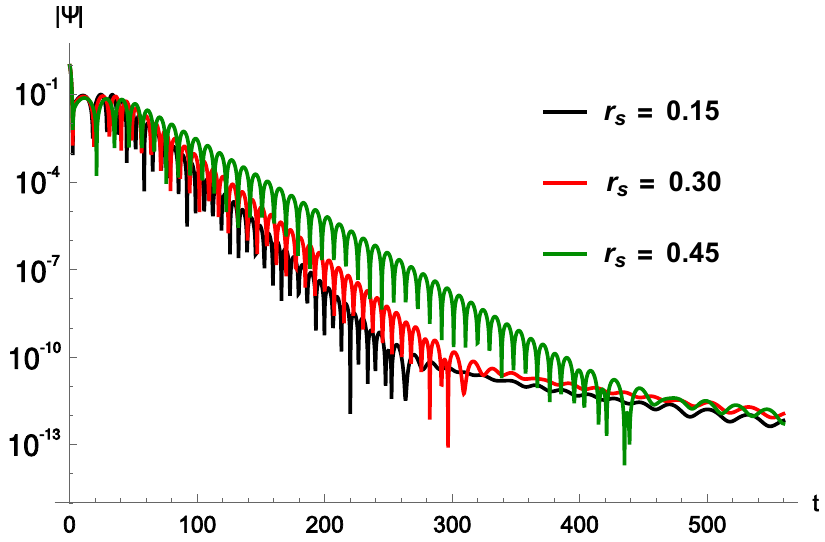}
    \end{minipage}
    \begin{minipage}[b]{0.32\textwidth}
        \centering
        \includegraphics[width=\textwidth]{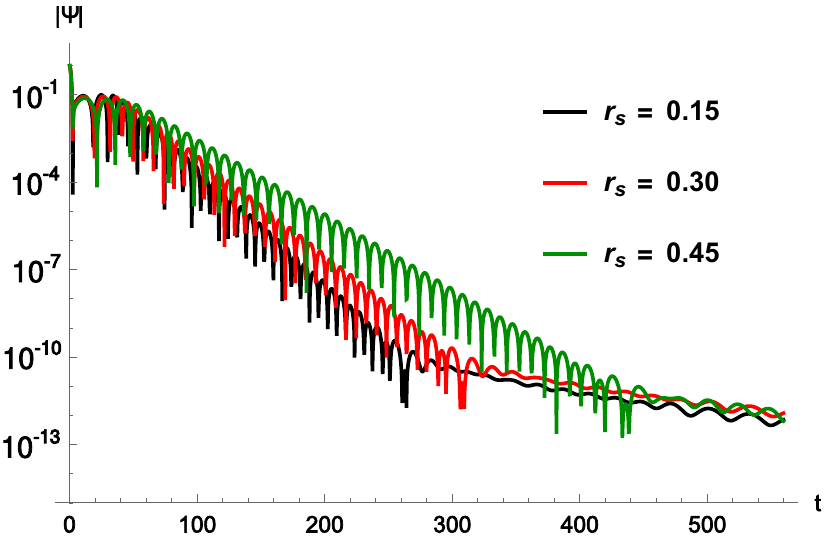}
    \end{minipage}%
    \begin{minipage}[b]{0.32\textwidth}
        \centering
        \includegraphics[width=\textwidth]{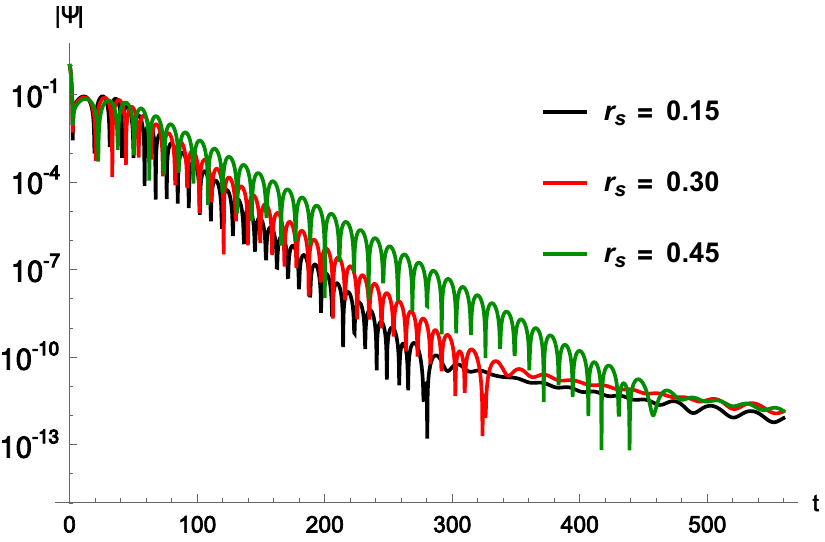}
    \end{minipage}
    \caption{The time-domain profiles from left to right correspond to the scalar field, electromagnetic field, and axial gravitational field  ($ \rho_s = 0.01 $, $ l = 2 $).}
    \label{fig:6}
\end{figure*}
\noindent factor, and $O(\delta^4)$ is a higher - order infinitesimal term, indicating that as the step size $\delta$ decreases, the influence of this part on the formula result becomes relatively smaller, which reflects the accuracy of the numerical approximation. In numerical calculations, explicit initial conditions are required to initiate the calculation process. 

 The initial perturbations in many physical scenarios exhibit characteristics similar to a Gaussian distribution\cite{Moderski:2001ru}. The initial condition of a Gaussian pulse provides a definite starting state for describing the field perturbation wave function, which can be written as:

\begin{equation}
\psi(\mu = u_0, \, v) = A \exp \left[ - \frac{(v - v_0)^2}{\sigma^2} \right]
\end{equation}
$A$ is the amplitude of the pulse, $v_0$ is the central position parameter of the Gaussian pulse, and $\sigma$ is the standard deviation, which reflects the dispersion degree of the pulse. Through the above - mentioned solution, the time evolution of the black hole's quasi - normal modes can be obtained. We can effectively extract information such as amplitude, phase, frequency, and damping factor from the data using the Prony method.

\begin{equation}
\psi(t) \simeq \sum_{i = 1}^{p} C_i e^{-i \omega_i t}
\end{equation}

\subsection{Numerical Results}
The time - domain profile is of central importance in the study of black hole quasi - normal modes. It directly reflects the dynamic behavior of spacetime perturbations during time evolution. As can be seen from Equations~\ref{equ:24} and~\ref{equ:27}, the effective potential is directly related to the quasinormal modes of the black hole. In Section 2, the spatial diagram of the effective potential has been drawn. Here, we will use the same parameter values to plot the time - evolution diagram of the waveform. In Figures~\ref{fig:5}, when $ \rho_s = 0 $, it corresponds to the black curve in the figure, which is the time - domain profile of the Schwarzschild black hole. It can be divided into three stages: the initial stage, the quasinormal mode stage, and the power - law tail stage. As shown in Figures~\ref{fig:5} and~\ref{fig:6}, regardless of how $ \rho_s $ and $ r_s $ change, they have a similar profile to the Schwarzschild black hole. This is because their effective potentials have the same asymptotic behavior. When $ \rho_s = 0 $ and $ r_s $ increase, the oscillation of the wave function slows down over time, the oscillation time increases, and the oscillation frequency decreases.

\begin{figure*}[t!]
  \includegraphics[width=0.48\textwidth]{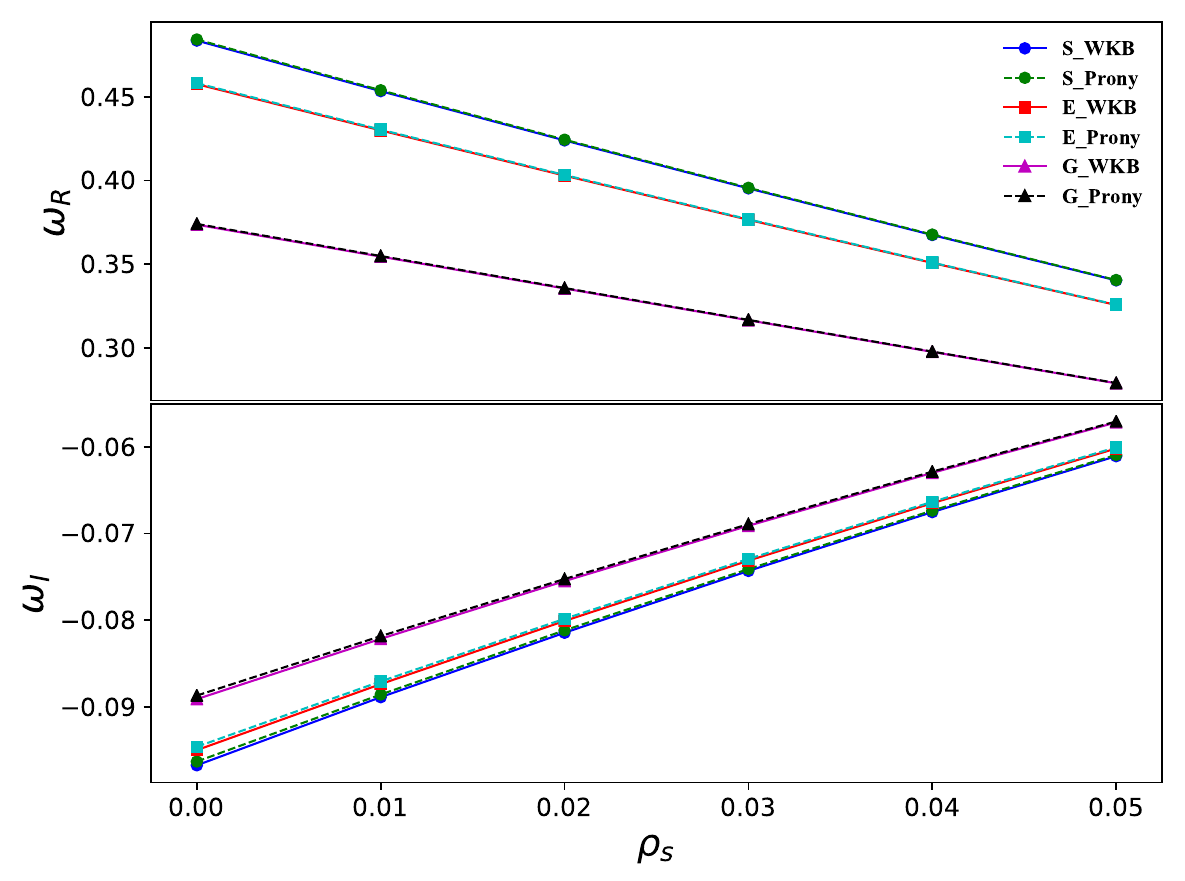}
    \includegraphics[width=0.48\textwidth]{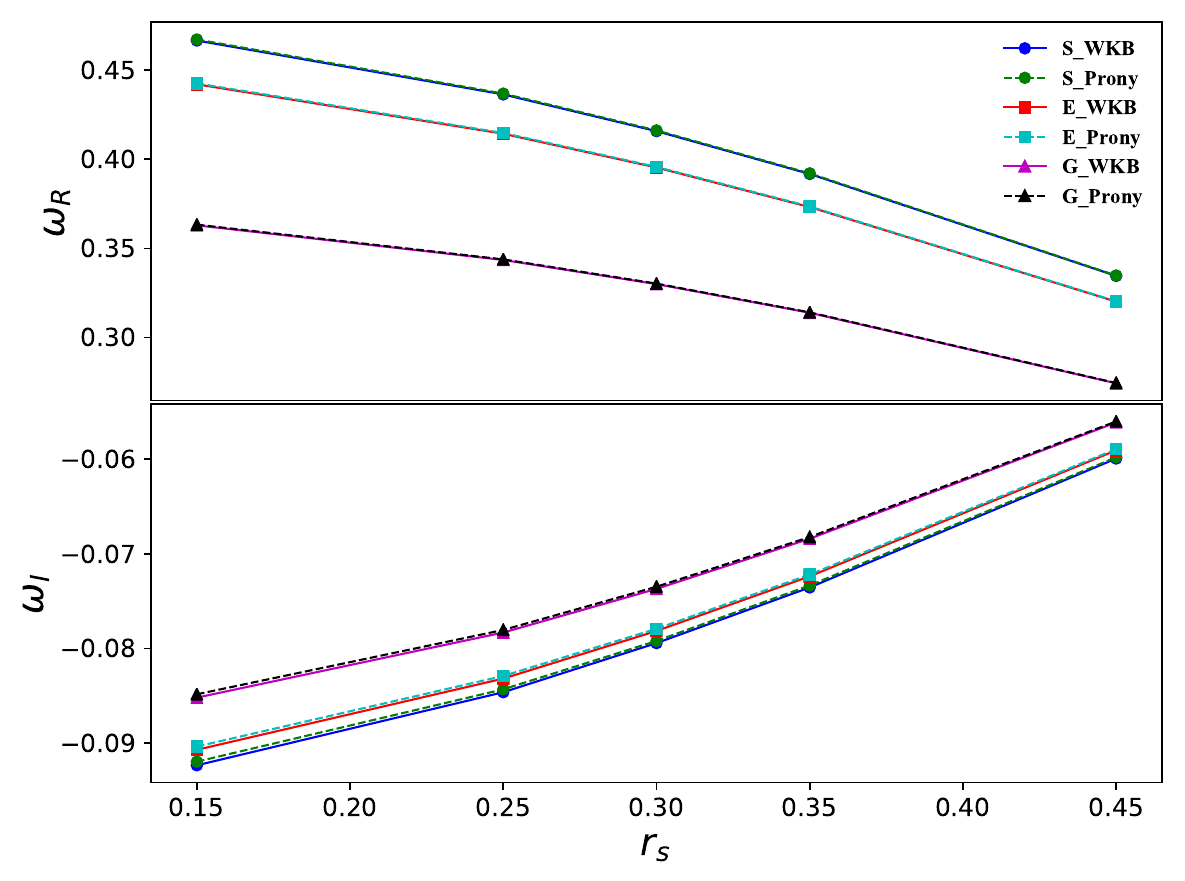}
\caption{The variations of the real and imaginary parts of the quasinormal mode frequencies under the perturbations of different fields with the characteristic density $\rho_s = 0.01$ (left) and the radius $r_s = 0.2$ (right)}
\label{fig:7}       
\end{figure*}

It is difficult to discern the frequencies of the quasinormal modes from Figure~\ref{fig:5} and~\ref{fig:6} as shown. Therefore, in Figure~\ref{fig:7}, we have plotted the real and imaginary parts of the complex frequencies of the quasinormal modes extracted by the WKB and Prony methods. In Figures~\ref{fig:7}, S represents the scalar field, E represents the electromagnetic field, and G represents the axial gravitational field. The legend applies to both the real and imaginary parts. We distinguish between the two methods by using shapes of the same type but in different colors, namely circles, squares, and triangles. From the figure, it can be seen that the shapes overlap very well, indicating a high degree of consistency in the frequencies extracted by the two methods. It also demonstrates that the extracted data has high reliability.  As the $r_s$ and density $\rho_s$ 

\begin{table*}[]
\caption{The frequencies of the quasinormal modes corresponding to the multipole quantum number $l$ under different field perturbations, $\rho_s = 0.03$ ,$r_s = 0.25$.}
\label{tab:2}       
\begin{tabular}{rcccccc}
\hline
     & \multicolumn{2}{c}{Scalar}   & \multicolumn{2}{c}{electromagnetic}  & \multicolumn{2}{c}{Axial gravitational}     \\
\hline
\multicolumn{1}{c}{$l$} & \multicolumn{1}{c}{WKB} & \multicolumn{1}{c}{Prony} & \multicolumn{1}{c}{WKB} & \multicolumn{1}{c}{Prony} & \multicolumn{1}{c}{WKB} & \multicolumn{1}{c}{Prony} \\
\hline
0     & 0.068567-0.068823i  & 0.071494-0.067636i        &   --          &       --          &       --     &      --               \\
1     & 0.20892-0.0634604i  & 0.208972-0.063351i        & 0.18271-0.060706i       & 0.182701-0.060607i        &     --                    &   --              \\
2     & 0.347855-0.062922i  & 0.348101-0.062746i        & 0.33257-0.0619921i      & 0.332799-0.061835i        & 0.283946-0.0588i        & 0.284112-0.058738i       \\

3     & 0.486878-0.062785i & 0.487499-0.062526i   & 0.476044-0.062308i  &  
0.476626-0.062070i   & 0.442563-0.060795i      & 0.443043-0.060585i       \\
\hline

\end{tabular}
\end{table*}
\noindent increase, the real part gradually decreases, which corresponds to the slowdown of the oscillation frequency in the time-domain profile plots. The negative imaginary part indicates that the perturbations of the black hole under these fields are stable. The decrease in the absolute value of the imaginary part implies that the attenuation effect of the system weakens. That is, the rate of energy loss of the system during the evolution process decreases, and the system can maintain a certain state for a longer time. With the changes in these two parameters, the oscillation and attenuation characteristics of the system change synergistically, suggesting that factors such as the spatial distribution of the dark matter halo and the gravitational field of the black hole are intertwined and influence the overall behavior of the system. This provides important clues for studying the physical mechanism of the interaction between the dark matter halo and the black hole.

Finally, we discussed the influence of the multipole quantum number $l$ on the complex frequency of the quasinormal modes, and the results are presented in Table \ref{tab:2}. The multipole quantum number $l$ satisfies $l \geq s$, where $s$ is the spin of the perturbation field. The spin numbers $m$ corresponding to the scalar field, electromagnetic field, and axial gravitational  are 0, 1, and 2, respectively. According to the data in the table, when the multipole quantum number $l$ increases, the real part of the complex frequency of the quasinormal modes rises significantly (for example, in the scalar field, it increases from 0.068567 to 0.48), which reflects the enhancement of energy quantization. The imaginary part fluctuates slightly around $0.06i$, indicating that the dissipation mechanism of the system is stable.

\section{Conclusions}
\label{sec:4}

This paper conducts a systematic study on the quasinormal modes of a Schwarzschild black hole in a Dehnen-(1,4, 5/2) type dark matter halo. Prior to the study, the data within the 3$\delta$ confidence interval of the shadow radius of the M87* black hole was used as a constraint range to determine the values of the Dehnen dark matter halo parameters $ r_s $ and  $ \rho_s$ , and it was clarified that there is a negative correlation between them. Subsequently, the wave equations and effective potentials of the black hole under scalar field, electromagnetic field, and axial gravitational perturbations were derived. The results show that the presence of the dark matter halo can change the effective potential of the black hole. The larger the values of $ r_s $ and  $ \rho_s$, the lower the peak value of the effective potential. This change reduces the energy "barrier" faced by particles when moving around the black hole, enhancing the particle activity. Meanwhile, the "resistance" of the system to external perturbations weakens, thus affecting the matter distribution and energy transfer around the black hole.

When exploring the quasinormal modes of the black hole, two numerical methods, namely the WKB method and the time - domain method, were utilized. The results show that as $ r_s $ and  $ \rho_s $ increases, the oscillation of the wave function slows down over time, the oscillation time lengthens, and the oscillation frequency decreases. In terms of the real and imaginary parts of the quasinormal mode frequencies, the real part gradually decreases, which is consistent with the phenomenon of a slower oscillation frequency in the time - domain profile. The imaginary part is negative, and its absolute value decreases, indicating that the black hole is stable under the perturbations of these fields. The energy dissipation rate is reduced, allowing the black hole to maintain a certain state for a longer time.In addition, the study also explored the influence of the multipole quantum number $l$ on the complex frequencies of the quasinormal modes. When $l$ increases, the real part of the quasinormal mode complex frequency rises significantly, reflecting enhanced energy quantization. The imaginary part fluctuates slightly around $0.06i$, indicating that the system's dissipation mechanism is stable.Our research provides relevant data on the quasinormal modes of Schwarzschild black holes in Dehnen - (1,4, 5/2) type dark matter halos, offering a reference for testing different dark matter - black hole models.

\section{Acknowledgements}
This research was  supported by the National Natural Science Foundation of China (Grant No.
12265007).

\bibliographystyle{apsrev4-2}



\end{document}